\documentclass[showpacs,aps,prb,twocolumn,groupedaddress]{revtex4}
\usepackage{graphicx} 
\usepackage[dvipdfm]{hyperref}
\hypersetup{colorlinks=true,linkcolor=blue,
  implicit=true,breaklinks=true,pagebackref=true,backref=true,
bookmarks=true,bookmarksnumbered=true,hyperfootnotes=true,debug=true,
  naturalnames=false,citecolor=blue,pdfview=FitH,pdfstartview=FitH,hyperindex=true}
\usepackage{amsmath}
\usepackage{amssymb}
\usepackage{ifsym}

\newcommand{\bk}{\mathbf k}

\newcommand{\beg}{\begin{equation}}
\newcommand{\en}{\end{equation}}
\newcommand{\dg}{^\dagger}

\begin{document}

\title{Pressure studies of the quantum critical alloy Ce$_{0.93}$Yb$_{0.07}$CoIn$_5$} 

\author{Y. P. Singh$^{1}\footnote{These authors have contributed equally to this work}$, D. J. Haney$^{1*}$, X. Y. Huang$^1$, B. D. White$^2$, M. B. Maple$^2$, M. Dzero$^1$
and C. C. Almasan$^1$}
\affiliation{$^1$Department of Physics, Kent State University, Kent, Ohio, 44242, USA, \\
$^2$ Department of Physics, University of California at San Diego, La Jolla, CA 92903, USA}

\date{\today}

\begin{abstract}
Here we present  our experimental and theoretical study of the effects of pressure on the transport properties of the heavy-fermion alloy Ce$_{1-x}$Yb$_{x}$CoIn$_{5}$ with actual concentration $x\approx0.07$.
We specifically choose this value of ytterbium concentration because the magnetic-field-induced quantum critical point, which separates the antiferromagnetic and paramagnetic states at zero temperature, approaches zero, as has been established in previous studies.
Our measurements show that pressure further suppresses quantum fluctuations in this alloy, just as it does in the parent compound CeCoIn$_5$.
In contrast, the square-root temperature dependent part of resistivity remains insensitive to pressure, indicating that the heavy-quasiparticles are not involved in the inelastic scattering processes leading to such a temperature dependent resistivity.
We demonstrate that the growth of the coherence temperature with pressure, as well as the decrease of the residual resistivity, can be accurately described by employing the coherent potential approximation for a disordered Kondo lattice. 

\end{abstract}

\maketitle

\section{Introduction}
Since their discovery almost thirteen years ago,\cite{Petrovic2001,Sarrao2007} the family of `115'  materials has provided an impactful experimental and theoretical playground for studying fundamental quantum phenomena, such as magnetism and superconductivity, in strongly interacting electronic systems.\cite{Coleman2007}
In particular, the physical and structural properties of these materials have not only helped to further develop the concepts of quantum phase transitions and non-Fermi liquids, but have also motivated theoretical studies of exotic mechanisms for unconventional superconductivity.
Moreover, it has been shown recently that $f$-orbital compounds  may host topologically non-trivial electronic states. \cite{Dzero2010,Hasan2010,Wolgast2013,Kim2013,Zhang2013,Kim2014}
Whether the `115'-based alloys can host topologically non-trivial superconductivity remains an open question, which provides an additional motivation for both experimental and theoretical communities to study the normal and superconducting properties of these systems in greater detail. 

Heavy-fermion alloys Ce$_{1-x}$Yb$_x$CoIn$_5$  - members of the `115' family of compounds - possess a number of intriguing and often counterintuitive physical properties:
(i) upon an increase in the concentration of ytterbium atoms, the critical temperature of the superconducting transition decreases only slightly compared to other rare-earth substitutions \cite{Shu2011,Paglione2007} and superconductivity persists up to the nominal concentration $x_{nom}\sim 0.75$;
(ii) the value of the out-of-plane magnetic field ($H$) corresponding to the quantum critical point (QCP) approaches zero as $x_{nom}\to 0.2$;\cite{Hu2013}
(iii) there is a crossover in the temperature ($T$) dependence of resistivity:
for $x_{nom}<0.2$, resistivity remains predominately linear in temperature,
while for $x_{nom}>0.2$, resistivity has a square-root temperature dependence: \cite{Singh2014}
\begin{equation}\label{Eq1}
\rho(x,T) = \rho_{0}(x) + A(x)T + B(x) \sqrt T
\end{equation}  
with $\rho_{0}(x)\propto x_{nom}(1-x_{nom})$ (in accord with Nordheim law), \cite{Onuki1987,Kappler1981}
$B(x)\to0$ as $x_{nom}\to0$ and $A(x_{nom})\to0$ as $x_{nom}$ is gradually increased from zero to $x_{nom}\simeq 0.2$;
(iv) there is a drastic Fermi-surface reconstruction for $x_{nom}\simeq 0.55$, yet the critical temperature ($T_c$) of the superconducting transition remains weakly affected. \cite{Polyakov2012}
More recently, penetration depth measurements \cite{Kim2_2014} have shown the disappearance of the nodes in the superconducting order parameter for $x_{nom}>0.2$.
Finally, we note that recent studies indicate that the nominal ytterbium concentration $x_{nom}$ is related to the actual concentration $x_{nom}\approx 3x_{act}$ for $x_{nom} \leq 0.4$.

The emergent physical picture which describes the physics of these alloys is based on the notion of co-existing electronic networks coupled to conduction electrons:
one is the network of cerium ions in a local moment regime, while the other consists of ytterbium ions in a strongly intermediate-valence regime.\cite{Verma1976,Stewart1984}
This picture is supported by recent extended x-ray absorption fine structure (EXAFS) spectroscopic measurements \cite{Booth2011}, as well as photoemission, x-ray absorption, and thermodynamic measurements. \cite{Dudy2013,White2012}
Moreover, our most recent transport studies \cite{Singh2014} are generally in agreement with this emerging physical picture.
In particular, for $x_{nom}\approx 0.6$ we observe the crossover from coherent Kondo lattice of Ce to coherent behavior of Yb sub-lattice, which is in agreement with recent measurements of the De Haas-van Alphen (dHvA) effect,\cite{Polyakov2012} while superconductivity still persists up to $x_{nom}\approx 0.75$ of ytterbium concentration.
Nevertheless, it remains unclear which of the conduction states - strongly or weakly hybridized - of the stoichiometric compound contribute to each network.

In order to get further insight into the physics of the Ce$_{1-x}$Yb$_x$CoIn$_5$ alloys, we study the transport properties under an applied magnetic field and pressure for the alloy with actual concentration $x\approx 0.07$.
One of our goals is to clarify the origin of the square-root temperature dependence of resistivity and to probe the contribution of the heavy-quasiparticles to the
value of $B(x)$ [see Eq.~(\ref{Eq1})]. To address this issue, we study the changes in the residual resistivity and the coefficients $A$ and $B$ [see Eq.~(\ref{Eq1})] with pressure.
Our results show that while both the residual resistivity and the coefficient $A$ decrease with pressure, $B$ shows very weak pressure dependence.
We find that the Kondo lattice coherence and the superconducting critical temperature increase with pressure in accord with general expectations.\cite{Singh2014,Zhang2002}
We also study theoretically the properties of a disordered Kondo lattice in which the disorder ions are ``magnetic''.
Within the picture of the single conduction band, we show that the presence of the magnetic ions has little effect on the dependence of the residual resistivity and the Kondo lattice coherence temperature on pressure.
Our theoretical results are in good agreement with our experimental findings. 

Another important aspect of the present work concerns the evolution of the physical quantities affected by the presence of the field-induced quantum critical point.
In our recent work, \cite{Hu2013,Singh2014} we have shown that the temperature dependence of the magnetic field $H_{max}$ at which magneto-resistivity has a maximum  is a signature of system's proximity to field-induced QCP.
Consequently, here we study the dependence of $H_{max}$ on pressure.
We find a remarkable similarity between the dependence of the residual resistivity and $(d{H}_{max}/dT)^{-1}$ on pressure.
Yet, this result is not surprising because it is well understood that the tendency towards antiferromagnetic ordering originates from the partial screening of the $f$-moments by conduction electrons. Hence, a strong pressure dependence of the relevant physical quantities such as $A$ and $H_{max}$ is expected.

This paper is organized as follows.
In the next Section we provide the details of our experimental measurements.
The results of our measurements are presented in Section~III.
Section~IV is devoted to theoretical modeling of a disordered Kondo lattice under pressure.
Specifically, we find that both the residual resistivity and the coefficient in front of the leading temperature-dependent term decrease under pressure, in agreement with our experimental results.
In Section~V we provide the discussion of our results and conclusions. 

\section{Experimental details}
Single crystals with an actual composition Ce$_{0.93}$Yb$_{0.07}$CoIn$_{5}$ were grown using an indium self-flux method.
The crystal structure was determined from X-ray powder diffraction, while the composition was determined according to the method developed by  Jang et al.\cite{Jang2014}
The single crystals have a typical size of $2.1 \times 1.0 \times 0.16$ mm$^3$, with the $c$-axis along the shortest dimension of the crystals.
They were etched in concentrated HCl for several hours to remove the indium left on the surface during the growth process and were then rinsed thoroughly in ethanol. Four leads were attached to the single crystals, with current $I \parallel a$-axis, using a silver-based conductive epoxy.
We performed in-plane resistivity and transverse ($H\perp ab$) magnetoresistance (MR) measurements  as a function of temperature between 2 and 300 K, applied magnetic field up to 14 T, and applied hydrostatic pressure ($P$) up to 8.7 kbar. 

\section{Experimental results}

Figure~1(a) shows the in-plane electrical resistance ($R_a$) data as a function of temperature of a Ce$_{0.93}$Yb$_{0.07}$CoIn$_{5}$ single crystal measured under pressure. The qualitative behavior of resistance is the same for all pressures used in this study: the resistance initially decreases as the sample is cooled from room temperature, then it passes through a minimum in the temperature range 150 K to 200 K, followed by an increase as the temperature is further lowered. This increase is logarithmic in $T$, in accordance to the single-ion Kondo effect. With the onset of coherence effects at the Kondo lattice coherence temperature ($T_{coh}$), the resistance decreases with decreasing temperature below $T_{coh}$, while at even lower $T$, superconductivity sets in at $T_c$.  
\begin{figure}
\centering
\includegraphics[width=1.0\linewidth]{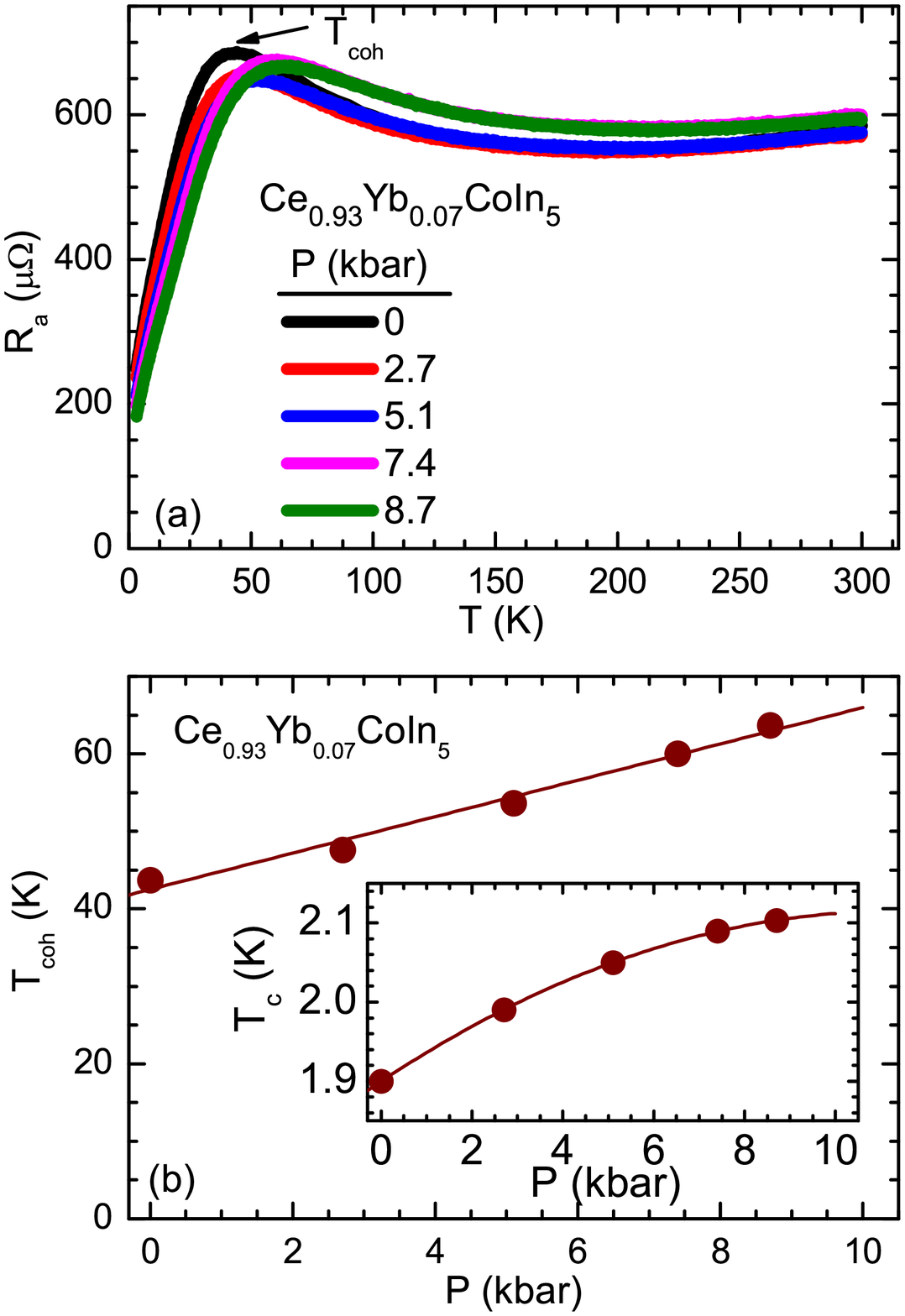}
\caption{(Color online) (a) Resistance $R_a$ of Ce$_{0.93}$Yb$_{0.07}$CoIn$_{5}$ as a function of temperature $T$ for different pressures $P$ (0 kbar, 2.7 kbar, 5.1 kbar, 7.4 kbar, 8.7 kbar). The arrow at the maximum of the resistance data represents the coherence temperature $T_{coh}$. (b) Evolution of $T_{coh}$ as a function of pressure $P$. Inset:  Superconducting critical temperature $T_c$ as a function of pressure $P$.  In each, the solid line is a guide to the eye.}
\label{Fig1}
\end{figure}

\begin{figure}
\centering
\includegraphics[width=1.0\linewidth]{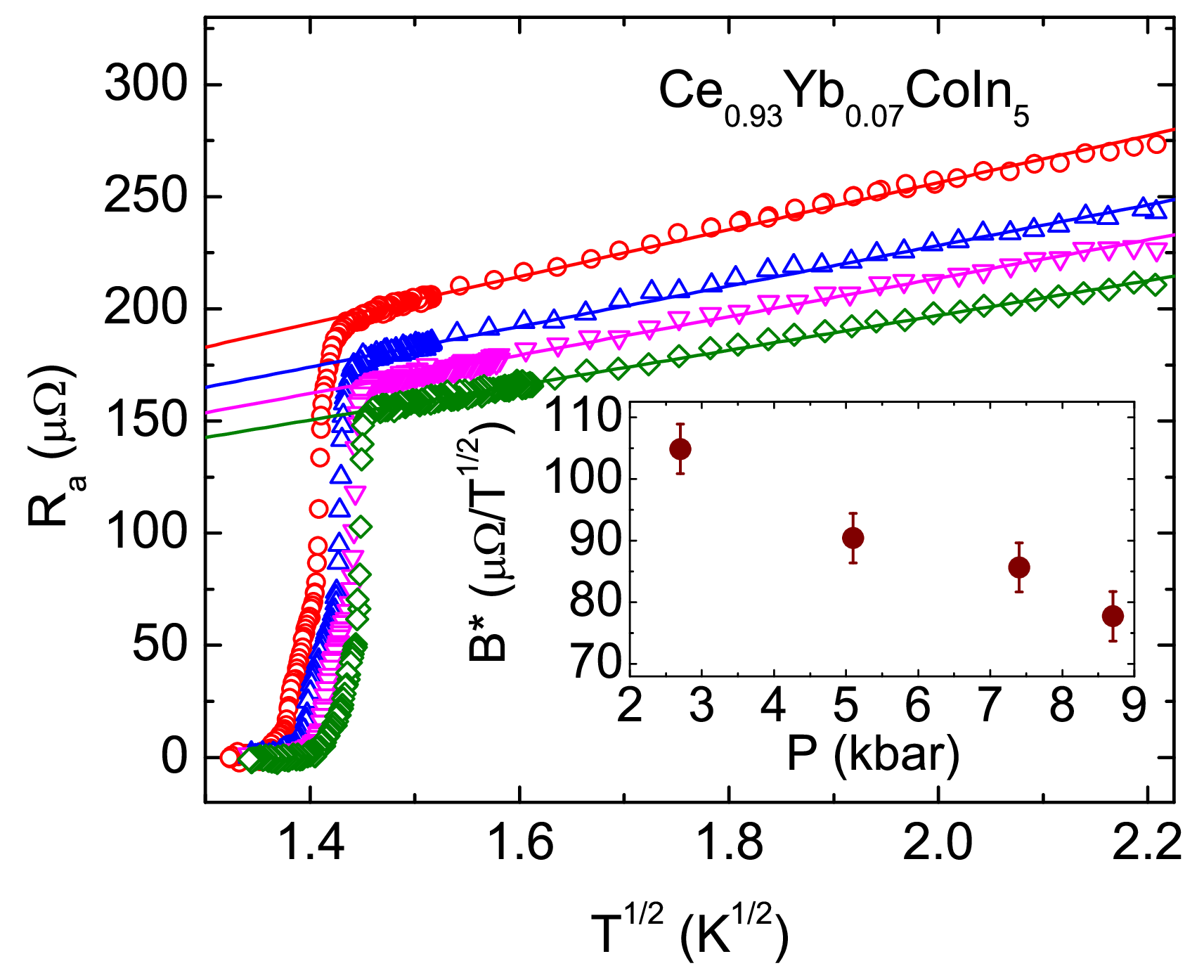}
\caption{(Color online) Resistance $R_a$ of Ce$_{0.93}$Yb$_{0.07}$CoIn$_{5}$ as a function of $\sqrt{T}$, in the temperature range $1.8\text{ K} \leq T \leq 5 \text{ K}$. The solid lines are linear fits to the data. Inset: Pressure $P$ dependence of coefficient $B^*$ of $\sqrt{T}$, obtained from fits of the resistance data shown in main panel.}
\label{Fig2}
\end{figure}

The onset of coherence is governed by the process in which the $f$-electrons of Ce can resonantly tunnel into the conduction band, i.e.,
$f^1\rightleftharpoons f^0+e$.
Because the cell volume $\Omega$ changes due to these resonant processes, i.e., $\Omega(f^1)-\Omega(f^0)>0$, the electronic properties are strongly susceptible to the application of external pressure.
Thus, we expect that pressure increases the local hybridization of Ce$_{0.93}$Yb$_{0.07}$CoIn$_{5}$ and, hence, increases the coherence temperature (see Section IV for a related discussion). Figure~1(b) shows that, indeed, the disordered Kondo lattice $T_{coh}$ increases with increasing pressure, just as it does for pure CeCoIn$_{5}$ and the other members of the Ce$_{1-x}$R$_x$CoIn$_{5}$ ($R=$ rare earth) series. \cite{White2012} The inset to Fig.~1(b) shows the pressure dependence of the superconducting critical temperature $T_c$.
An increase in $T_c$ with pressure is expected because at low temperatures, $T_{coh}$ sets an effective bandwidth, so that $T_c \propto T_{coh}$. 

 It is well known 
\cite{Koitzsch2013,Maehira2003,Settai2001,Barzykin2007} that large and small Fermi surfaces co-exist in the stochiometric CeCoIn$_5$. 
An open question is: Do the electrons from the small Fermi surface hybridize with ytterbium ions or only the electrons from the large Fermi surface
hybridize with both cerium and ytterbium ions? The former (latter) scenario would give a pressure independent (dependent) coefficient for the temperature dependence of the scattering processes. Therefore, to address this question, we study the changes in the temperature-dependent part
of resistivity under pressure. 

As we have already discussed in the Introduction, we have previously shown that there are two distinct contributions to the scattering of the quasi-particles in Ce$_{1-x}$Yb$_x$CoIn$_{5}$ alloys:
a linear in $T$ contribution due to quantum critical fluctuations, observed for smaller Yb doping ($x_{nom} \leq0.2$), and a $\sqrt{T}$ contribution, which emerges at higher $x_{nom}$ [see Eq. (1)]. In what follows we trace out the changes in the coefficients $A$ and $B$ with pressure for the Ce$_{0.93}$Yb$_{0.07}$CoIn$_{5}$ alloy, for which both of these contributions are present at least over a certain temperature range and under ambient pressure.  

Figure~2 shows the $R_a$ data vs $\sqrt{T}$ around the superconducting transition temperature ($1.8 \leq T \leq 5$ K). This figure shows that from just above $T_c$ to 5 K the $R_a(T)$ data follow very well the $\sqrt T$ dependence (solid lines are linear fits to the data). The pressure dependence of the coefficient $B^*$ in front of $\sqrt{T}$, obtained from the fitting of these data, is plotted in the inset to Fig.~2. Notice that there is a significant suppression  of $B^*$ as pressure increases. This pressure dependence of $B^*$ suggests that the scattering just above $T_c$ is largely governed by fluctuating Cooper pairs originating from the heavy Fermi surface. This observation is in agreement with the fluctuation correction to resistivity due to pre-formed Cooper pairs composed of heavy quasiparticles. Indeed, for a  three-dimensional (3D) Fermi surface and in the case of a strong coupling superconductor with relatively small coherence length \cite{DeBeer2006}, one expects a $\sqrt{T}$ fluctuation contribution to resistivity.\cite{Tinkham2012} Therefore, these $R_a(T)$ data show that the linear in $T$ contribution of Eq. (1), due to the system's proximity to the field-induced QCP, is masked by the strong SC fluctuations near the SC boundary.

\begin{figure}
\centering
\includegraphics[width=1.0\linewidth]{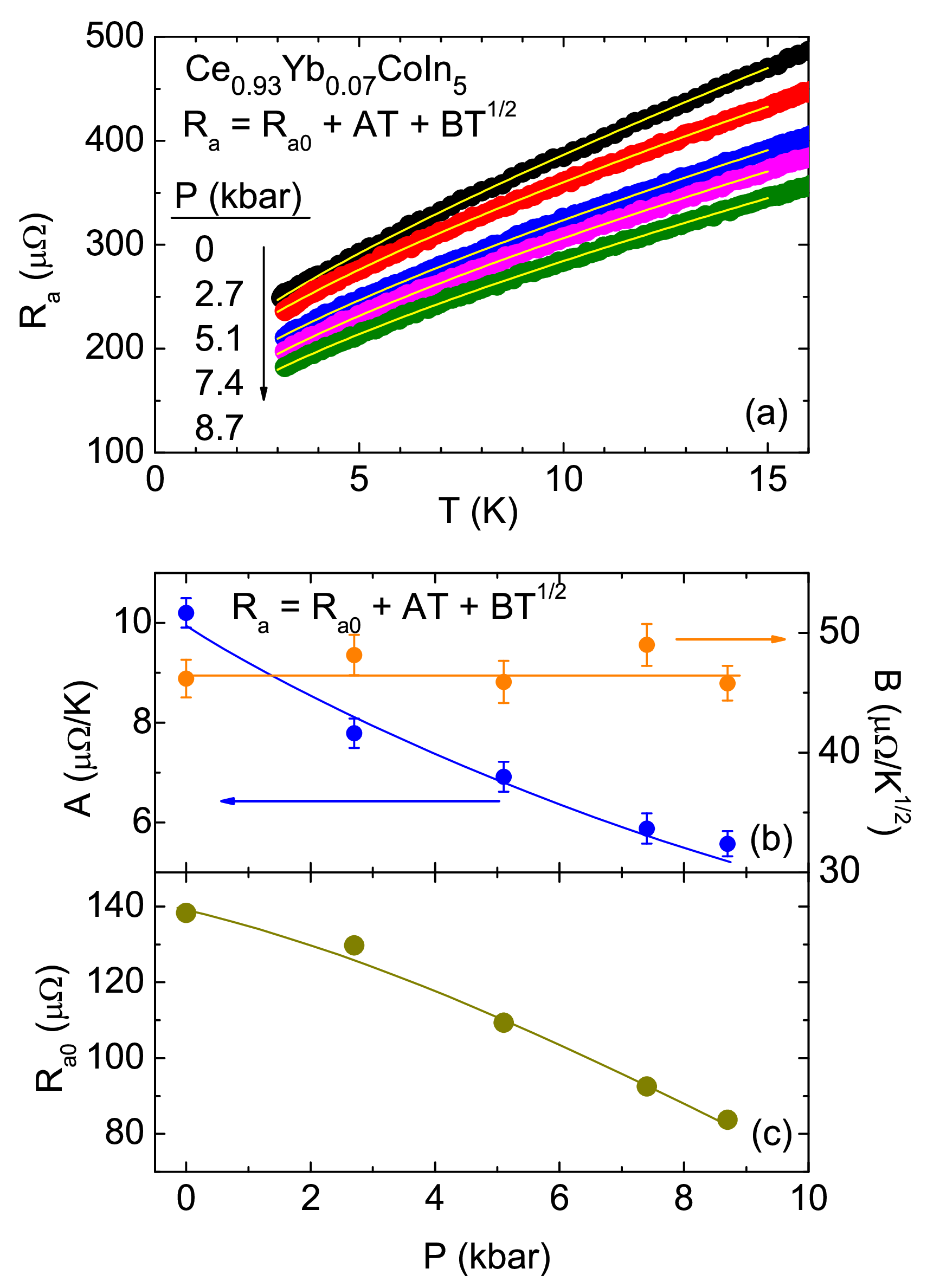}
\caption{(Color online) (a) Fits of the resistance $R_a$ data with $R_a(p,x,T) = R_{a0}(p,x) + A(p,x)T + B(p,x) \sqrt{T}$ for different pressures for Ce$_{0.93}$Yb$_{0.07}$CoIn$_{5}$ in the temperature range $3 \text{ K} \leq T \leq 15 \text{ K}$. (b) Pressure $P$ dependence of the linear $T$ contribution $A$ and $\sqrt{T}$ contribution $B$, obtained from fits of the resistance data shown in panel~(a). (c) Pressure $P$ dependence of the residual resistance R$_{a0}$, obtained from the fits.}
\label{Fig3}
\end{figure}  

\begin{figure}
\centering
\includegraphics[width=1.0\linewidth]{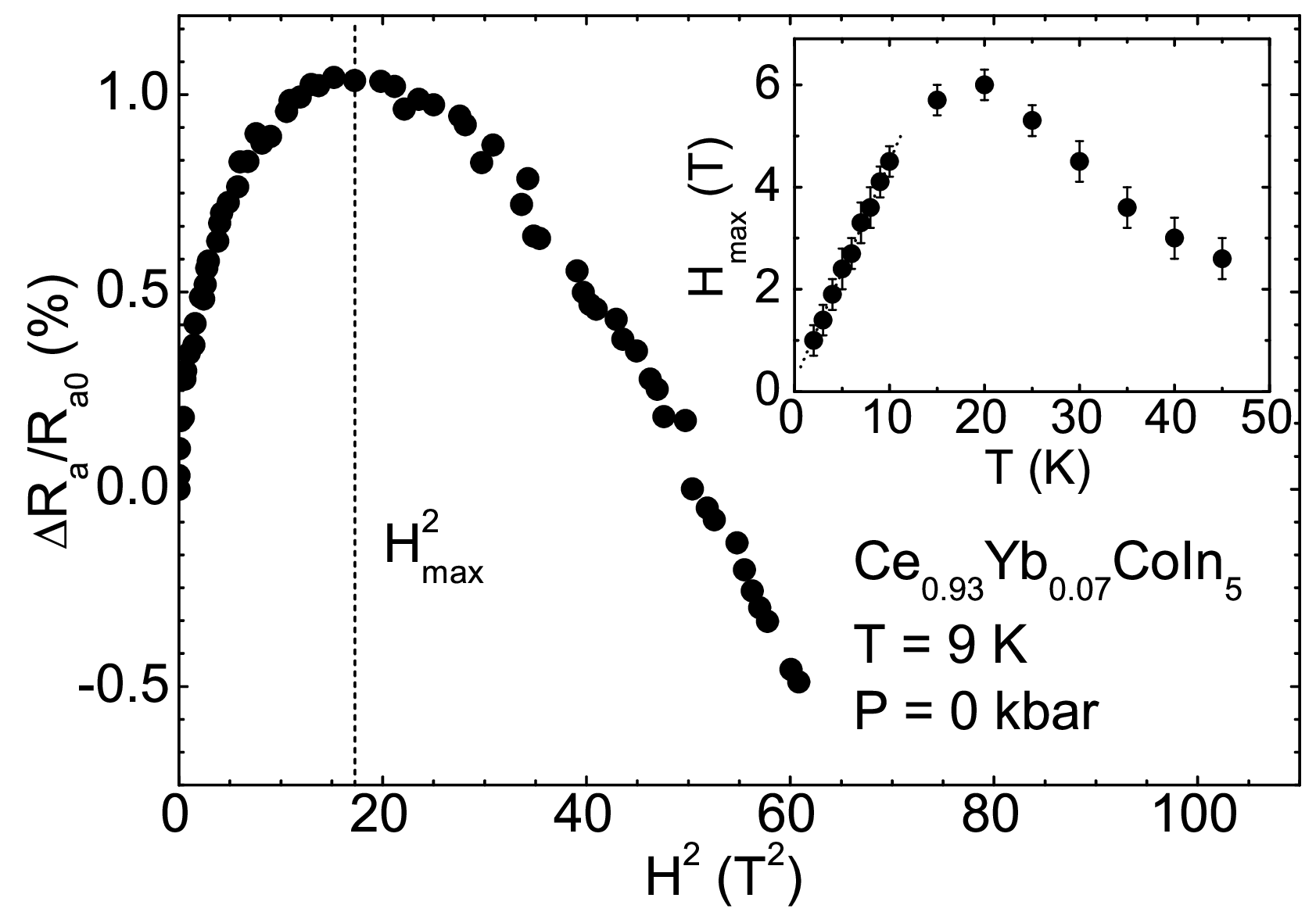}
\caption{(Color online) (a) Magnetic field $H$ dependence (plotted as function of $H^2$) of electrical resistance $R_a$ normalized to its zero field value for Ce$_{0.93}$Yb$_{0.07}$CoIn$_{5}$. The dashed line in the main figure marks $H_{max}$, corresponding to the coherence giving way to single-ion Kondo behavior. Inset: $H_{max}$ vs temperature $T$.} 
\label{Fig4}
\end{figure}

The superconducting fluctuations, nevertheless, decrease  as the system moves away from $T_c$ to higher temperatures. Indeed, the resistivity data starts deviating from the $\sqrt{T}$ behavior for temperatures above $T\approx 5$ K. In fact, as shown in Fig.~3(a), the data are fitted very well with Eq. (1) (the solid lines are the fits) for $3 \leq T \leq 15$ K and for all pressures studied. From these fits we obtain the pressure dependence of the fitting parameters $R_{a0}$, $A$, and $B$, which allows us to probe the relative contribution of heavy- and light-quasiparticle states to scattering. 

Figure 3(b) shows the pressure dependence of the parameters $A$ and $B$ extracted from the fitting of $R_a(T)$ of Fig. 3(a), which, as discussed above, are the weights of the linear-in-T and square-root-in-T scattering dependences, respectively. Notice that $A$ decreases while  $B$ remains relatively constant with increasing pressure. The suppression of $A$ with pressure indicates that quantum fluctuations are suppressed under increasing pressure. Also, the insensitivity of $B$ to pressure suggests that  the inelastic scattering events leading to the $\sqrt T$ dependence in this temperature range involve quasiparticles from the small Fermi surface. Hence, these $R_a(T)$ data for $T \geq 5$ K show that the strong SC fluctuations of the heavy quasiparticles that give the $\sqrt T$ dependence just above $T_c$ are not anymore the dominant scattering contribution at these higher temperatures and that the dominant contribution to scattering comes now from the AFM quantum fluctuations of the heavy quasiparticles (with a linear in $T$ scattering behavior) and the quasiparticles from the small Fermi surface (with a $\sqrt T$ scattering behavior). 

Moreover, the increase in the values of the parameter $B(P=0,x)$ with increasing ytterbium concentration\cite{Hu2013} is likely due to an increase in the size of the initially small Fermi pockets near the $M$ points of the 3D Brillouin zone. In this context, a recently observed disappearance of the nodes in the superconducting order parameter for $x_{nom} \approx 0.2$,\cite {Kim2_2014} although seen in other heavy-fermion compounds,\cite{Huang2014} is particularly intriguing given the fact that in the parent compound the order parameter has $d_{x^2-y^2}$ symmetry.\cite{Park2008,Park2009}
Consequently, vanishing of the nodes suggests that for $x_{nom}>0.2$  the order parameter may have exotic symmetry, which in principle can give rise to topologically protected surface states.\cite{Qi2011,Dzero2014}
However, to verify the realization of specific scenarios for the symmetry of superconducting order parameter in Ce$_{1-x}$Yb$_x$CoIn$_5$, one would need a detailed understanding of the electronic properties in both normal and superconducting states \cite{Dzero2014}. 

Figure 3(c) shows the pressure dependence of residual resistance R$_0$ extracted from the fitting of the data of Fig. 3(a). Since tuning with pressure does not introduce any impurity scattering in the system, the decrease in residual resistance with increasing pressure indicates again that the scattering due to AFM quantum spin fluctuations is suppressed by pressure. Indeed,  quantum fluctuations in this family of heavy fermion superconductors are known to be suppressed by pressure because the magnetic order in the Ce-lattice is suppressed.\cite{Jaccard1999,Grosche1996,Hu2_2012}

Next, we present the results of the magnetoresistance (MR) measurements on the $x=0.07$ ($x_{nom}= 0.2$) single crystals.
We performed the in-plane ($I \parallel a$-axis) transverse ($H\perp I$) MR measurements, defined as $\Delta R_a/R_{a0} \equiv [R_a(H)-R_a(0)]/R_a(0)]$, on Ce$_{0.93}$Yb$_{0.07}$CoIn$_{5}$ in applied magnetic fields up to 14~T and for temperatures ranging from 2 to 50 K.
The main panel of Fig. 4 shows one such MR curve  measured at 9 K, which is typical for this whole temperature range.
The MR shows non-monotonic $H$ dependence: it decreases quadratically with field at high fields, typical of a single-ion Kondo system, while it increases with field at low fields, for up to $H=H_{max}$. This latter positive MR is due to the formation of the coherent Kondo lattice state.
$H_{max}$ represents the value where the coherent state gives way to the single-ion state due to the fact that magnetic field breaks the coherence of the Kondo lattice. \cite{Coleridge1987,Ruvalds1988,Flouquet1988,Rauchschwalbe1987,Aronson1989,Ohkawa1990}

In a conventional Kondo lattice system, as $T$ increases, $H_{max}$ moves toward lower field values, signifying that a lower field value is sufficient to break coherence at these higher temperatures due to thermal fluctuations, with a complete suppression of the positive contribution to MR, hence $H_{max}=0$, at $T \approx T_{coh}$. On the other hand, as we had recently revealed,\cite{Hu2013} $H_{max}(T)$ in the Ce$_{1-x}$Yb$_x$CoIn$_5$ alloys with concentrations $x_{nom} \leq 0.2$  shows deviation from the conventional Kondo behavior and exhibits a peak, below which $H_{max}$ decreases with decreasing temperature. This is shown for Ce$_{0.93}$Yb$_{0.07}$CoIn$_{5}$ in the inset of Fig. 4, which is a plot of the temperature dependence of $H_{max}$ at ambient pressure. We have attributed the decrease  in $H_{max}(T)$ with decreasing $T$ to quantum spin fluctuations that dominate the MR behavior below about 20 K.\cite{Hu2013}
$H_{max}(T)$ shows linear behavior below 10 K (see inset to Fig. 4).  A linear extrapolation of this low $T$ behavior to zero temperature gives $H_{QCP}$.\cite{Hu2013} Notice that $H_{QCP} \approx 0.2$ T in Ce$_{0.93}$Yb$_{0.07}$CoIn$_{5}$ at ambient pressure, as previously reported,\cite{Hu2013} showing that this Yb doping ($x = 0.07$) is close to the quantum critical value $x_c$ for the Ce$_{1-x}$Yb$_{x}$CoIn$_{5}$ alloys.

\begin{figure}
\centering
\includegraphics[width=1.0\linewidth]{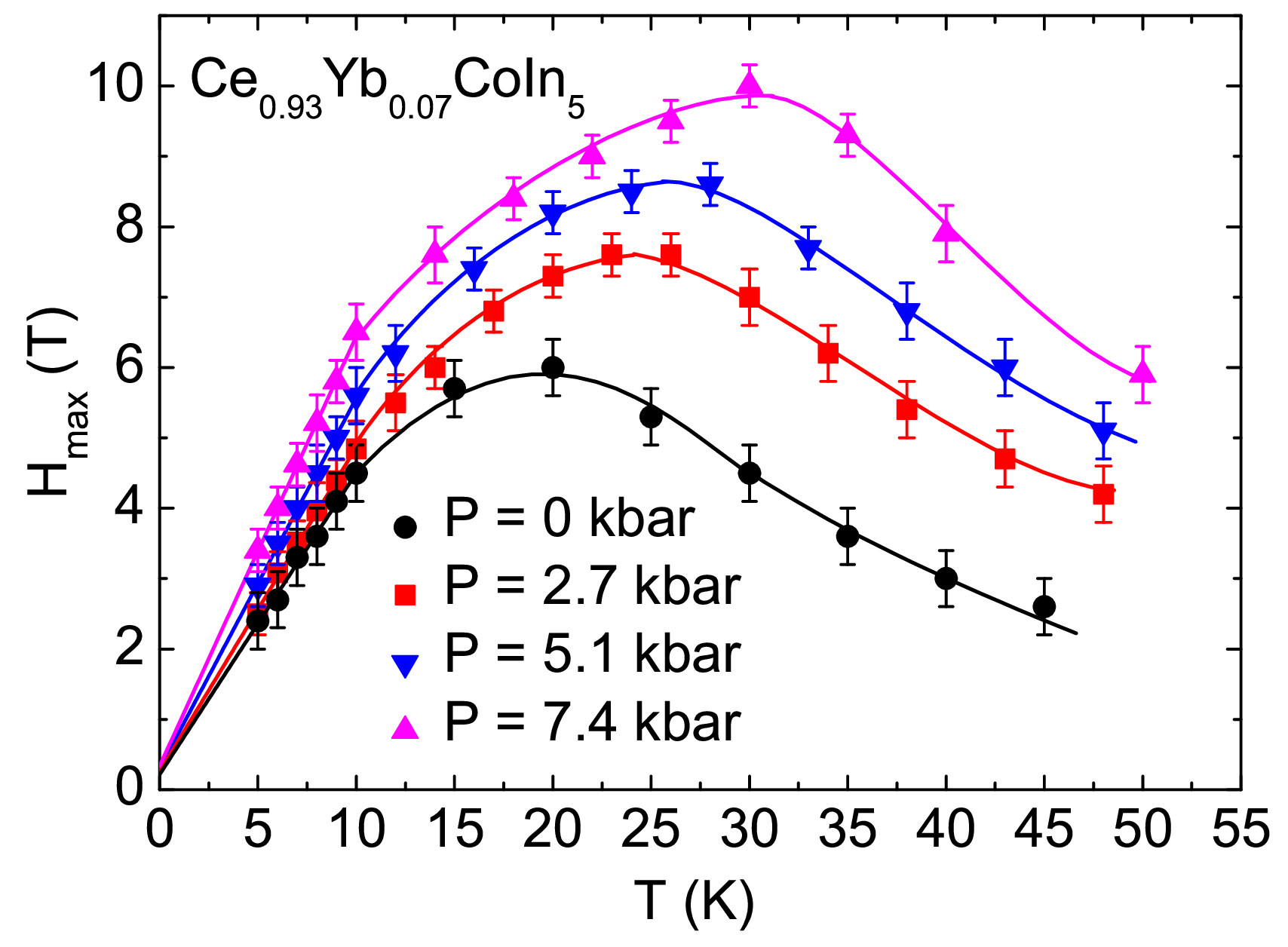}
\caption{(Color online) Temperature $T$ dependence of the maximum in magneto-resistance $H_{max}$ for different pressures $P$. The solid lines below $10$~K are linear fits to the data, while the other lines above $10$~K are guides to the eye.}
\label{Fig5}
\end{figure}
 
\begin{figure}
\centering
\includegraphics[width=1.0\linewidth]{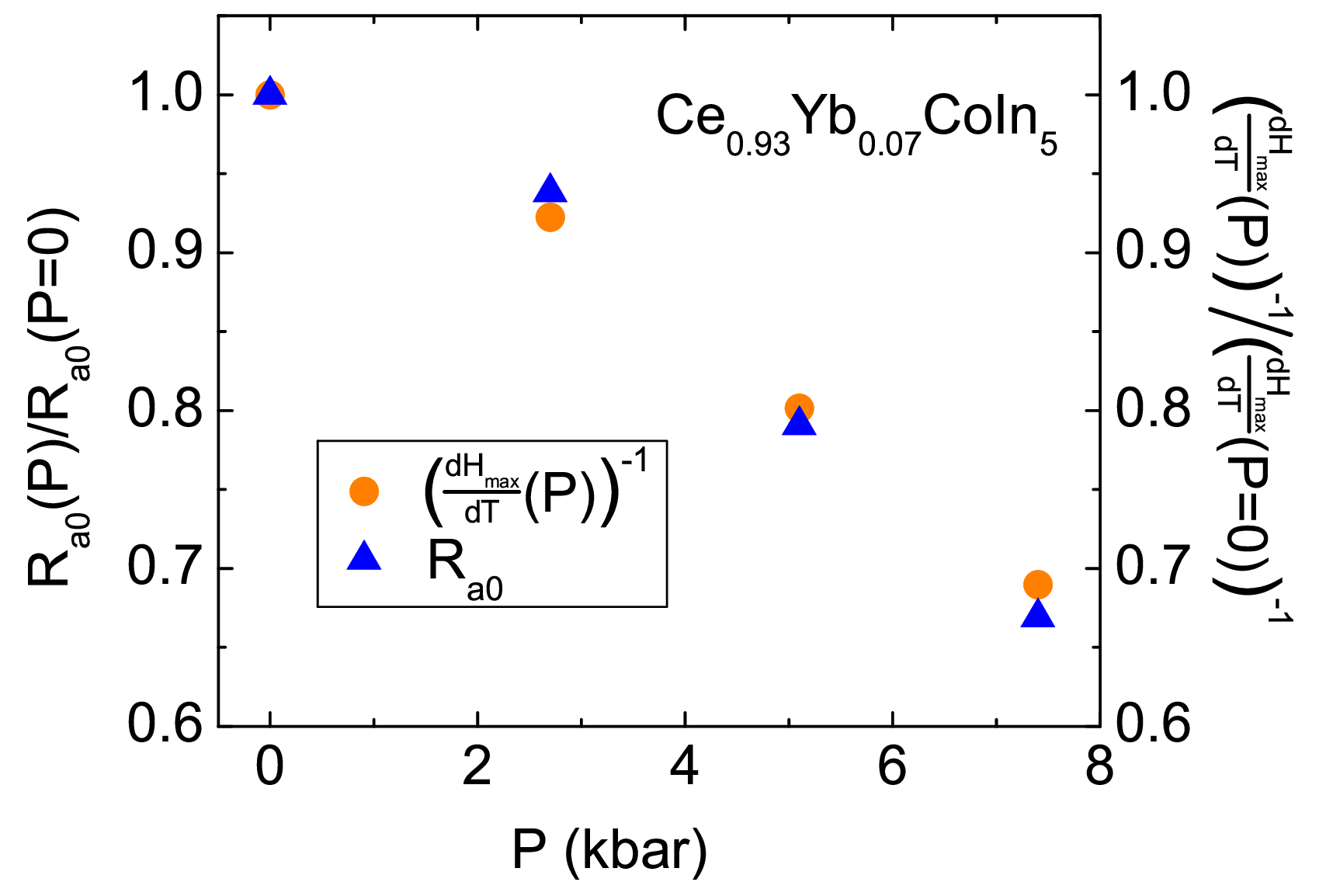}
\caption{(Color online) Pressure $P$ dependence of residual resistance $R_a0$ (obtained through the fitting of the resistance data as discussed in the text), normalized to its value at zero pressure (right vertical axis) and $P$ dependence of inverse slope of $H_{max}(T)$ normalized to its value at zero pressure (left vertical axis).}
\label{Fig6}
\end{figure}  

In order to better understand the quantum critical behavior in Yb-doped CeCoIn$_{5}$, we performed these same magnetoresistance measurements under pressure. Figure 5 shows the temperature dependence of $H_{max}$, extracted as discussed above from the MR data plotted as a function of magnetic field for four different applied hydrostatic pressures. Three notable features are revealed by this figure: (i) the application of pressure does not change qualitatively the $H_{max}(T)$ dependence, (ii) there is no noticeable change in the value of $H_{QCP}$ with pressure for $P \leq 8.7$ kbar, most likely because of the already small value of $H_{QCP}$ ($H_{QCP}=0.2$ T) at ambient pressure, and (iii) both the value of $H_{max}$ and the position in $T$ of the $H_{max}(T)$ peak shifts to higher temperatures with increasing pressure; as a result, the slope $dH_{max}/dT$ for $T < 10$ K increases with pressure. According to the Doniach phase diagram, \cite{Doniach1977} the Kondo temperature $T_K$ and the magnetic exchange interaction temperature $T_{RKKY}$ of Ce Kondo lattices increase with increasing pressure. Hence, the increase in $H_{max}$ with pressure is a result of increased $T_{coh}$, and the shift in the peak of $H_{max}(T)$ to higher $T$ with pressure is a result of the increase of both $T_{RKKY}$ and $T_{coh}$ with pressure.   
The increase in the slope $dH_{max}/dT$ with increasing pressure means that a larger applied field is required to break the Kondo singlet. We note that both quantum spin fluctuations and applied magnetic field contribute to the braking of Kondo coherence at temperatures $T<10$ K. Therefore, a larger $dH_{max}/dT$ at higher pressures can be understood in terms of weaker quantum spin fluctuations since a larger field is required to break the Kondo singlet compared with the field required for smaller $dH_{max}/dT$ where spin fluctuations are stronger.            

We show in  Fig. 6 the inverse of this slope as a function of pressure, normalized to its zero pressure value.
We also show in the same figure (right vertical axis) the residual resistance as a function of pressure, also normalized to its zero pressure value.
Notice that these two quantities scale very well, indicating that the same physics, i.e. the suppression of quantum critical fluctuations with pressure, dominate their behavior with pressure. 

%
%

\section{Theory}
In this Section, we will formulate a general approach to Kondo alloys diluted with magnetic dopants that will help us to interpret our experimental results.
In what follows, we first introduce the model in order to study the effects of pressure in disordered Kondo lattice.
Then, we will employ the coherent potential within the mean-field theory for the disordered Kondo lattice to compute the pressure dependence of the Kondo lattice coherence temperature and residual conductivity.
\subsection{Model}
We consider the following model Hamiltonian, which we write as a sum of three terms 
\beg\label{H}
\hat{\cal H}=\hat{H}_0+\hat{H}_{Kh}+\hat{H}_{V}\,.
\en
The first term describes the kinetic energy of the conduction and $f$-electrons in the unperturbed (i.e. spatially homogeneous) Kondo lattice:
\beg\label{H0}
\hat{H}_0=\sum\limits_{\bk\sigma}\epsilon_\bk \hat{c}_{\bk\sigma}\dg\hat{c}_{\bk\sigma}+\sum\limits_{\bk\sigma}\varepsilon_f\hat{f}_{\bk\sigma}\dg\hat{f}_{\bk\sigma}\,,
\en
where $\epsilon_\bk=-(t_c/2)(\cos k_x+\cos k_y)-\mu_c$ is the single particle energy taken relative to the chemical potential $\mu_c$ (here we will ignore the transport along the $z$-axis).
The second term in Eq. (\ref{H}) accounts for the Kondo holes, i.e., it prohibits the $f$-electrons from occupying an impurity site, and it also describes the impurity $f$-electrons denoted by $\hat{p}$:
\beg\label{Hh}
\begin{split}
\hat{H}_{Kh}&=\sum\limits_{i\sigma}(1-\xi_i)(\varepsilon_{0f}+\varepsilon_f)\hat{f}_{i\sigma}\dg\hat{f}_{i\sigma}+\sum\limits_{\sigma}\tilde{\epsilon}_f\hat{p}_\sigma\dg \hat{p}_\sigma+\\ &+
\frac{U_f}{2}\sum\limits_{i\sigma}\xi_i\hat{f}_{i\uparrow}\dg\hat{f}_{i\uparrow}\hat{f}_{i\downarrow}\dg\hat{f}_{i\downarrow}+
{U_p}\hat{p}_{\uparrow}\dg\hat{p}_{\uparrow}\hat{p}_{\downarrow}\dg\hat{p}_{\downarrow}\,,
\end{split}
\en
where summation goes over all lattice cites, and
\beg
\xi_i=\left\{\begin{matrix} 0, \quad i= 0 \\ 1, \quad i\not= 0\end{matrix}
\right.\,,
\en
with $i=0$ denoting the position of an impurity site.
The first term in Eq. (\ref{Hh}) accounts for an $f$-electron state on an impurity site.
Physically, this process cannot happen.
Therefore, at the end of the calculation, the energy of the $f$-electron on the impurity site will be taken to infinity, $\varepsilon_{0f}\to\infty$, to ensure $\langle \hat{f}_{i=0\sigma}\dg\hat{f}_{i=0\sigma}\rangle=0$.
Lastly, the third term in Eq. (\ref{H}) accounts for the hybridization between the conduction electrons and both cerium $f$-electrons and ytterbium $f$-holes:
\beg\label{Hv}
\hat{H}_V=\sum\limits_{i\sigma}\xi_i\left(V_f\hat{c}_{i\sigma}\dg\hat{f}_{i\sigma}+h.c.\right)+
\sum\limits_{\bk\sigma}\left(V_p\hat{c}_{\bk\sigma}\dg\hat{p}_{\sigma}+h.c.\right).
\en

Clearly, the theoretical analysis of this model is hindered by the presence of the Hubbard interaction terms with both $U_{f}$ and $U_p$ being the largest energy scales in the problem.
To make progress, we will adopt the slave-boson mean-field theory (SBMF) approach.
Thus, we will set $U_{f}$ and $U_p$ to infinity:
\beg
U_f\to\infty\,, \quad U_p\to\infty\,.
\en
The double occupancy on the $f$-sites is excluded by introducing the slave-boson projection operators:
\beg\label{project}
\begin{split}
&\hat{f}_{i\sigma}\to \hat{b}_i\dg\hat{f}_{i\sigma}\,, \quad \hat{f}_{i\sigma}\dg\to \hat{f}_{i\sigma}\dg\hat{b}_i\,, \\
&\hat{p}_{\sigma}\to \hat{a}\dg\hat{p}_{\sigma}\,, \quad \hat{p}_{\sigma}\dg\to \hat{p}_{\sigma}\dg\hat{a}\,,
\end{split}
\en
supplemented by the following constraint conditions:
\beg
\begin{split}
&\sum\limits_\sigma \hat{f}_{i\sigma}\dg\hat{f}_{i\sigma}+\hat{b}_i\dg\hat{b}_i=1, ~\sum\limits_\sigma \hat{p}_{\sigma}\dg\hat{p}_{\sigma}+\hat{a}\dg\hat{a}=1\,.
\end{split}
\en

Thus, the phase space is reduced to the set of either singly occupied states $|b^0f^1\rangle$ or empty states $|b^1f^0\rangle$ for the $f$-electrons and, similarly, $|a^0p^1\rangle$ or $|a^1p^0\rangle$ for $f$-holes.
Clearly, the hybridization part of the Hamiltonian in Eq. (\ref{Hv}) always acts only between these two states.
Thus, for the kinetic energy terms, we find
\beg
\hat{f}_{i\sigma}\dg\hat{f}_{i\sigma}|b^0f^1\rangle\to\hat{f}_{i\sigma}\dg\hat{b}_i\hat{b}_i\dg\hat{f}_{i\sigma}|b^0f^1\rangle=
\hat{f}_{i\sigma}\dg\hat{f}_{i\sigma}|b^0f^1\rangle\,.
\en

In the mean-field approximation, the projection (slave-boson) operators are replaced with their expectation values:
\beg
\hat{b}_i\to\langle\hat{b}_i\rangle=b\,, \quad \hat{a}\to\langle\hat{a}\rangle=a\,.
\en
The corresponding mean-field Hamiltonian is
\begin{widetext}
\beg\label{Hmf}
\begin{split}
\hat{\cal H}_{mf}&=\sum\limits_{\bk\sigma}\epsilon_\bk \hat{c}_{\bk\sigma}\dg\hat{c}_{\bk\sigma}+
\sum\limits_{\bk\sigma}\varepsilon_f\hat{f}_{\bk\sigma}\dg\hat{f}_{\bk\sigma}+\sum\limits_{i\sigma}(1-\xi_i)(\varepsilon_{0f}-\varepsilon_f)\hat{f}_{i\sigma}\dg\hat{f}_{i\sigma}+\sum\limits_{\sigma}\tilde{\epsilon}_f\hat{p}_\sigma\dg \hat{p}_\sigma+\sum\limits_{i\sigma}\xi_i\left(V_fb^*\hat{c}_{i\sigma}\dg\hat{f}_{i\sigma}+h.c.\right)+\\ & +
\sum\limits_{\bk\sigma}\left(V_pa^*\hat{c}_{\bk\sigma}\dg \hat{p}_{\sigma}+h.c.\right)+
\sum\limits_{i}\xi_i\lambda_b\left(\sum\limits_\sigma \hat{f}_{i\sigma}\dg\hat{f}_{i\sigma}+|b|^2-1\right)+\lambda_a\left(\sum\limits_\sigma \hat{p}_{\sigma}\dg\hat{p}_{\sigma}+|a|^2-1\right),
\end{split}
\en
\end{widetext}
where $\lambda_{a,b}$ are Lagrange multipliers, which will be computed self-consistently.
Let us introduce the following parameters:
\beg\label{params}
E_f=\lambda_b+\varepsilon_f\,, \quad E_{0f}=\varepsilon_{0f}-E_f\,, \quad {\epsilon}_f=\tilde{\epsilon}_f+\lambda_a\,.
\en
In addition, I introduce $z=1-x$ with $x$ being the concentration of Yb ions:
\beg
z=\frac{1}{N_s}\sum\limits_{i}\xi_i\,.
\en
In this expression $N_s$ is the total number of sites.
After re-arranging the terms in Eq. (\ref{Hmf}) and using Eq. (\ref{params}) we obtain:
\beg\label{Hmf2}
\begin{split}
\hat{\cal H}_{mf}&=\hat{H}_{mf}^{(b)}+\hat{H}_{mf}^{(a)}\,, \\
\hat{H}_{mf}^{(b)}&=\sum\limits_{\bk\sigma}\epsilon_\bk \hat{c}_{\bk\sigma}\dg\hat{c}_{\bk\sigma}+\sum\limits_{\bk\sigma}E_f\hat{f}_{\bk\sigma}\dg\hat{f}_{\bk\sigma}+E_{0f}\hat{f}_{0\sigma}\dg\hat{f}_{0\sigma}\\&+
\sum\limits_{i\sigma}\xi_i\left(V_fb^*\hat{c}_{i\sigma}\dg\hat{f}_{i\sigma}+
b\hat{f}_{i\sigma}\dg\hat{c}_{i\sigma}\right)+zN_s\lambda_b\left(|b|^2-1\right)\,, \\
\hat{H}_{mf}^{(a)}&=\sum\limits_{\sigma}{\epsilon}_f\hat{p}_\sigma\dg \hat{p}_\sigma+
V_p\sum\limits_{\bk\sigma}\left(a^*\hat{c}_{\bk\sigma}\dg \hat{p}_{\sigma}+a\hat{p}_{\sigma}\dg\hat{c}_{\bk\sigma}\right)\\&+\lambda_a\left(|a|^2-1\right)\,.
\end{split}
\en

Because ytterbium ions are in the mixed valence state, the hybridization amplitude $V_p\ll V_f$.
Moreover, we assume that the condensation temperature $T_{Yb}$ for the bosons $a$ is significantly smaller than the Ce Kondo lattice coherence temperature $T_{coh}$.
This assumption is justified by the similarity in the physical properties of the Yb ion in Yb$_{x}$Y$_{1-x}$InCu$_4$ and in Ce$_{1-x}$Yb$_x$CoIn$_5$:
the ytterbium valence state is close to Yb$^{3+}$ for $x_{nom}\ll 0.1$ and becomes Yb$^{2.5+}$ for $x_{nom}\sim 0.1$.
At the same time, in Yb$_{x}$Y$_{1-x}$InCu$_4$, for small $x$, the single site Kondo temperature is approximately $2$~K.\cite{Ocko2003}
Thus, in our choice of the bare model parameters, we must keep in mind that the condensation temperature for the $a$-bosons is lower than the one for the $b$-bosons, $T_{Yb}<T_{coh}$. 

\subsection{Coherent Potential Approximation}
To analyze the transport properties of the disordered Kondo lattice, we employ the coherent potential approximation (CPA).\cite{Velicky1968,Brouers1972,Li1988,Li1991,Zhang2002}
The idea of the CPA is to introduce an effective medium potential, which allows for an equivalent description of the disordered system.
In particular, the effective potential is considered to be purely dynamical. This approximation is valid when the scattering events on different impurity sites are independent. 
\begin{widetext}
To formulate the CPA, we introduce the Lagrangian for the disordered Kondo lattice (which is related to $\hat{H}_{mf}^{(b)}$):
\beg\label{eq:LagrangeMF}
\begin{split}
{\cal L}=&\sum\limits_{\bk\sigma}\left[\hat{c}_{\bk\sigma}\dg
\left({\partial_\tau}+\epsilon_\bk\right)\hat{c}_{\bk\sigma}+
\hat{f}_{\bk\sigma}\dg\left({\partial_\tau}+E_f\right)\hat{f}_{\bk\sigma}\right]+
\sum\limits_\sigma\hat{f}_{0\sigma}\dg(\partial_\tau+E_f)\hat{f}_{0\sigma}+zN_s\lambda_b\left(|b|^2-1\right)\\
+&\sum\limits_{i\sigma}\xi_i\left(V_fb^*\hat{c}_{i\sigma}\dg\hat{f}_{i\sigma}+
b\hat{f}_{i\sigma}\dg\hat{c}_{i\sigma}\right),
\end{split}
\en
where, for brevity, we omit the dependence of the fermionic fields on Matsubara time $\tau$.
Note that we have not included the terms that involve $p$-fermions. The reason is that the $p$-fermions can be formally integrated out, which will lead to the appearance of the self-energy correction $\Sigma_a(\tau-\tau')$ in the first term of Eq. (\ref{eq:LagrangeMF}).
However, to keep our expressions compact, we will include this term later when we analyze the transport properties. 
Within the frame of the CPA, we introduce an effective medium Lagrangian for the disordered Kondo lattice system as follows:
\beg\label{eq:Lmedium}
\begin{split}
{\cal L}_{eff}=&\int\limits_{0}^{\beta}d\tau'\sum\limits_{\bk\sigma}\hat{\psi}_{\bk\sigma}\dg(\tau)\left[
\begin{matrix}
\delta(\tau-\tau')\left({\partial_\tau}+\epsilon_\bk\right) + S_{cc}(\tau-\tau',z) & S_{cf}(\tau-\tau',z) \\ 
S_{fc}(\tau-\tau',z) & \delta(\tau-\tau')\left(\partial_\tau+E_f\right)+S_{ff}(\tau-\tau',z) 
\end{matrix}
\right]\hat{\psi}_{\bk\sigma}(\tau')\\&+zN_s\lambda_b\left(|b|^2-1\right),
\end{split}
\en
\end{widetext}
where $\beta=1/k_BT$, we introduced the two-component spinor $\hat{\psi}_{\bk\sigma}\dg=(\hat{c}_{\bk\sigma}\dg ~\hat{f}_{\bk\sigma}\dg)$ for brevity, and $S_{ab}(\tau,z)$ are the components of the coherent potential that we will have to determine self-consistently.
The self-consistency condition for the components of $S_{ab}(\tau,z)$ is obtained by requiring that the corresponding correlation functions for the effective Lagrangian, Eq. (\ref{eq:Lmedium}), are equal to the disorder-averaged correlators for the disordered Kondo lattice, Eq. (\ref{eq:LagrangeMF}).\cite{Velicky1968}
In the ``Kondo hole'' limit ($E_{0f}\to\infty$), it follows:
\beg\label{eq:Swx}
\hat{S}(i\omega_n,z)=\left(\begin{matrix} 0 & bV_f \\ b^*V_f & S_{ff}(i\omega_n,z) \end{matrix}\right),
\en
where $i\omega_n=\pi T(2n+1)$ is a fermionic Matsubara frequency and 
\beg\label{Fff}
\begin{split}
&S_{ff}(\omega,x){F_{ff}(\omega)}={z-1}\,, \\
&F_{ff}(\omega)\\&=\sum\limits_{\bk}\frac{\omega-\epsilon_\bk}{(\omega-\epsilon_\bk)(\omega-E_f-S_{ff}(\omega,z))-V_f^2|b|^2}\,.
\end{split}
\en
These equations allow us to compute the remaining component of the coherent potential (\ref{eq:Swx}).
$S_{ff}(i\omega,z)$ is a function of parameters $E_f$ and $b$, which will have to be computed self-consistently by minimizing the free energy.

\begin{figure}
\centering
\includegraphics[width=1.0\linewidth]{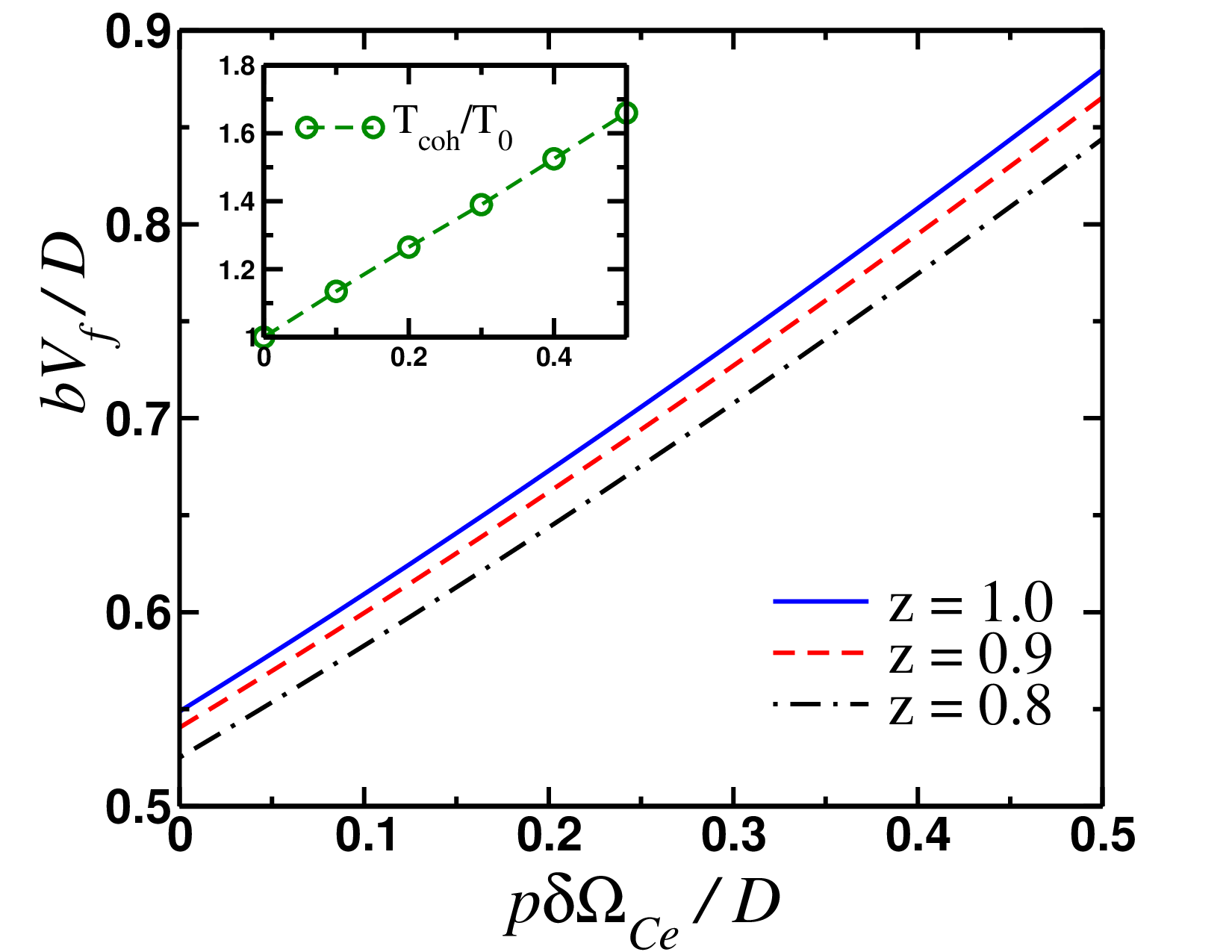}
\caption{(Color online) 
Pressure $p$ dependence of the slave-boson amplitude and coherence temperature $T_{coh}$ (inset) for various concentrations $z$ of the impurity $f$-sites.
The dependence of the coherence temperature $T_{coh}$ on pressure for $z=0.93$ is shown.}
\label{Fig7}
\end{figure}

\subsection{Slave Boson Mean-Field Theory for Disordered Kondo Lattice under Hydrostatic Pressure}
In order to study the effects of pressure in a disordered Kondo lattice, we need to express the change in the total volume of the system with the corresponding changes in the valence states of Ce and Yb ions.
For the Ce ions, the change in the $f$-shell occupation is positive due to its electronic nature, so that the resonance scattering involves a zero-energy boson, with amplitude $b$, and an electron: $f^{n+1}(j,m)\rightleftharpoons f^{n}(j,m)+e^-$.
In contrast, for the Yb ions, the resonance scattering involves a zero energy boson, with amplitude $a$, and a hole: $f^{n-1}(j,m)\rightleftharpoons f^{n}(j,m)+e^+$.
Thus, for the total volume of the system within the slave-boson mean-field theory, we write:\cite{Zhang2002}
\beg\label{eq:totvol}
\begin{split}
\Omega_t=(1-z)&[\Omega_{0Yb}+(1-a^2)\delta\Omega_{Yb}]\\+z&[\Omega_{0Ce}+(1-b^2)\delta\Omega_{Ce}]\,,
\end{split}
\en
where $\Omega_{0Yb,Ce}$ are the cell volumes for the singlet (non-magnetic) states on Yb ($f^{14}$) and Ce ($f^{0}$) ions, correspondingly.
Moreover, $\delta\Omega_{Yb,Ce}$ account for the difference in cell volumes between two $f$-ion configurations. 
Note that $\delta\Omega_{Yb}<0$ while $\delta\Omega_{Ce}>0$. 

To obtain the self-consistency equations for the slave-boson amplitude $b$ and constraint variable $\lambda_b$, we define the grand canonical enthalpy for an alloy under pressure $P$:
\beg\label{enthalpy}
\begin{split}
&K=-k_BT\log Z_{eff}\,, \\
&Z_{eff}=\textrm{Tr}\left\{e^{-\int\limits_{0}^\beta d\tau {\cal L}_{eff}(\tau)-P\Omega_t}\right\}.
\end{split}
\en
Minimizing the enthalpy with respect to $b$ and $\lambda_b$, we obtain:
\beg\label{eq:sb}
\begin{split}
&z\left(b^2-1\right)+2T\sum\limits_{i\omega_n}F_{ff}(i\omega_n)=0\,,\\
&zb(\lambda_b - P\delta\Omega_{Ce})+{2V_f}T\sum\limits_{i\omega_n}F_{fc}(i\omega_n)=0\,, 
\end{split}
\en
where $i\omega_n=i\pi T(2n+1)$ are Matsubara frequencies and
\beg
\begin{split}
&F_{fc}(\omega)=bV_f\\ &\times\sum\limits_{\bk}\frac{1}{(\omega-\epsilon_\bk)(z-E_f-S_{ff}(\omega,z))-
V_f^2|b|^2}\,.
\end{split}
\en
In addition, the third equation is the conservation of the total number of particles $N_{tot}=n_c+zn_f$, with
\beg\label{Fcc}
\begin{split}
&n_c=T\sum\limits_{i\omega_n}\sum\limits_{\bk}e^{i\omega_n0+}G_{cc}(\bk,i\omega_n)\,, \\
&G_{cc}(\bk,\omega)\\&=\frac{\omega-E_f-S_{ff}(\omega,z)}{(\omega-\epsilon_\bk)(\omega-E_f-S_{ff}(\omega,z))-
V_f^2|b|^2-\frac{V_p^2a^2}{\omega-\epsilon_f}}\,,
\end{split}
\en
which allows us to determine the renormalized position of the chemical potential $\mu_c$.
We note that equations that determine the value of $a$ and $\lambda_a$ can be obtained in the same manner as the ones above.
 
As a result, we find that the slave-boson amplitude $b$ grows linearly with pressure,\cite{Zhang2002} $b\propto P\delta\Omega_t$, see Fig~\ref{Fig7}. 
Also, our analysis
of the mean-field equations (\ref{eq:sb}) in the limit $b\to 0$ shows that the Kondo lattice coherence temperature $T_{coh}$ also grows with 
pressure almost linearly (Fig.~\ref{Fig7}):
\beg
T_{coh}\simeq E_f(T_{coh})\propto P\delta\Omega_t\,,
\en
which is in agreement with our experimental observations [see Fig. Fig1(b)].
In addition, as expected, we find that
(i) both slave-boson amplitude and coherence temperature decrease as the concentration of ytterbium atoms increases, and
(ii) the presence of the ytterbium $f$-electrons leads to a small reduction in the value of $b(P)$ relative to the case when $a=0$. 

\subsection{Transport Properties}
\begin{figure}
\centering
\includegraphics[width=0.8\linewidth]{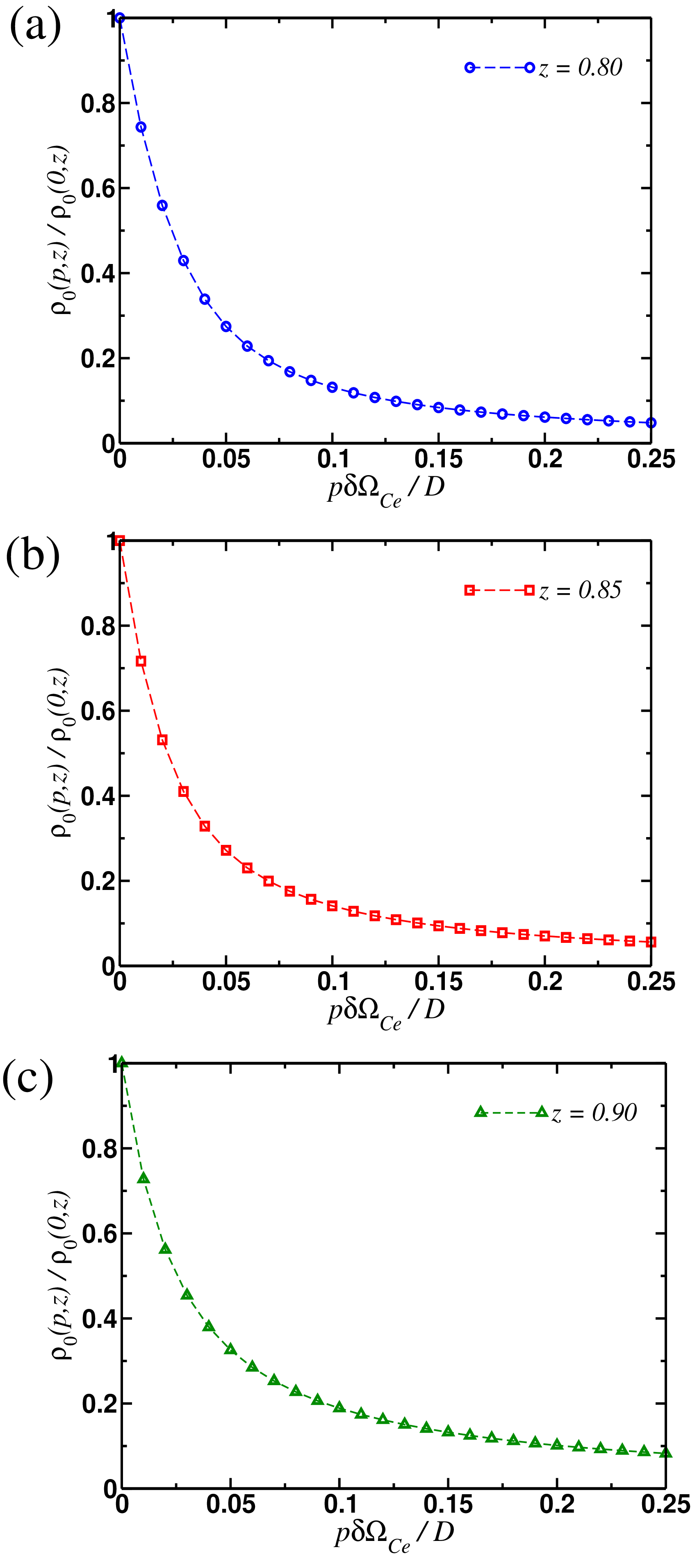}
\caption{(Color online) 
Pressure $p$ dependence of the residual resistivity $\rho_0$ for various alloy concentrations $z$.}
\label{Fig8}
\end{figure}

In this subsection we discuss the pressure dependence of the residual resistivity of the disordered Kondo lattice described by the Hamiltonian
(\ref{Hmf}).
We compute conductivity using the following expression: \cite{Mahan2000}
\beg\label{sigma}
\sigma_{\alpha\beta}(i\Omega)=\frac{1}{\Omega}\left\{\Pi_{\alpha\beta}(i\Omega)-\Pi_{\alpha\beta}(0)\right\},
\en
where $\alpha,\beta=x,y$, $s_\alpha=\sin k_\alpha$, $v_F$ is a Fermi velocity of the heavy-quasiparticles, and
\beg
\begin{split}
&\Pi_{\alpha\beta}(i\Omega)=e^2v_F^2T\sum\limits_{i\omega_n}\\&\times\sum\limits_\bk s_\alpha G_{cc}(\bk,i\omega_n+i\Omega)s_\beta G_{cc}(\bk,i\omega_n).
\end{split}
\en
To obtain the dependence of conductivity on the real frequency, we will perform the analytic continuation from $\Omega_n=2\pi Tn>0$ to real frequencies $i\Omega_n\to\omega$.
The residual resistivity can be computed from $\rho_0=\sigma^{-1}(\omega\to 0)$.
We present our results on Fig.~\ref{Fig8}.
In agreement with our experimental results, we find that the residual resistivity decreases with pressure.
At ambient pressure, the residual resistivity grows linearly with ytterbium concentration, which is again expected given our CPA approximation. 

The temperature dependence of resistivity can also be obtained from Eq. (\ref{sigma}).
Naturally, we find a `square-T' dependence: $\rho(P,T;z)=\rho_0(P,z)+A_{FL}(P,z)T^2$.
Because $A_{FL}(P,z)$ decreases with pressure, as does the coefficient in front of the linear-in-$T$ term in Eq. (\ref{Eq1}), we conclude that the inelastic scattering of heavy-quasiparticles determines the value of $A(P,z)$.

\section{Conclusions}
In this paper, we studied the Ce$_{0.93}$Yb$_{0.07}$CoIn$_{5}$ alloy ($x_{nom}=0.2$) using transport and magneto-transport measurements under hydrostatic pressure.
Our resistivity data reveal that the scattering close to $T_c$ follow a $\sqrt T$ dependence, consistent with the composite pairing theory,\cite{Erten2014} which for a three-dimensional (3D) system gives a $\sqrt{T}$ contribution to resistivity close to $T_c$, with a coefficient that decreases with increasing pressure. This latter result implies that the scattering in this $T$ range is largely governed by the heavy-quasiparticles from the heavy Fermi surface, hence it may reflect the scattering of composite pairs \cite{Erten2014} as a result of superconducting fluctuations. At higher $T$, our data reveal the presence of two scattering mechanisms: one linear in $T$ with a coefficient $A$ that decreases with increasing pressure and the other one with a  $\sqrt T$ dependence with a coefficient $B$ that is pressure independent. Given that the strong pressure dependence of the $A$ parameter directly relates to the strongly hybridized conduction and cerium $f$-electron states, we believe that the linear temperature dependence of the resistivity is governed by the scattering of heavy-quasiparticles,
while the scattering processes leading to the $\sqrt{T}$-term in resistivity are governed by the scattering of light electrons from the small Fermi surface.
Since the linear $T$ dependence is a result of quantum spin fluctuations, the decrease of $A$ with increasing pressure implies that quantum fluctuations are suppressed with pressure. This conclusion is confirmed by the fact that residual resistivity also decreases with pressure.

We also performed magnetoresistance measurements under applied hydrostatic pressure in order to study the evolution of quantum critical spin fluctuations with pressure. First, our magnetoresistance data reveal that the Ce$_{0.93}$Yb$_{0.07}$CoIn$_{5}$ alloy for this Yb doping ($x = 0.07$) is close to the quantum critical value $x_c$ for the Ce$_{1-x}$Yb$_{x}$CoIn$_{5}$ alloys. Second, these data confirm our findings from resistivity measurements that quantum critical fluctuations are suppressed with increasing pressure. Finally, we also analyzed the temperature and pressure dependence of the magnetic field $H_{max}$ at which magnetoresistance reaches its maximum values. At low temperatures, $H_{max}$ grows linearly with temperature. 
Interestingly, we find that the slope $d{H}_{max}/dT$ also grows with applied pressure, similarly to the dependence on pressure of the coherence temperature. These result suggests that the magneto-resistance is largely governed by the heavy-electrons from the large Fermi surface.  

Our theoretical analysis of the disordered Kondo lattice model with ``magnetic'' disorder ions shows that despite the presence of ``magnetic'' impurities rather than ``Kondo holes'', the coherence temperature grows and residual resistivity decreases with pressure as expected for ``electron-like'' Kondo ions.\cite{Zhang2002}
The growth of the coherence temperature leads to the corresponding growth of the superconducting critical temperature, indicating that superconductivity originates predominantly from the ``heavy'' Fermi surface.

\section{Acknowledgments} 

MD thanks Instituto Superior
T\'{e}chnico (Lisboa, Portugal), where part of this work has been completed, for hospitality.
This work was supported by the National Science Foundation (grant NSF DMR-1006606) and Ohio Board of Regents (grant OBR-RIP-220573) at KSU, and by the U.S. Department of Energy (grant DE-FG02-04ER46105) at UCSD. 

\bibliography{Ref20Yb}

\end{document}